\documentclass[3p,twocolumn,final,times]{elsarticle}
\usepackage{amssymb}
\usepackage{lineno}
\usepackage{makecell}
\usepackage{ulem}
\usepackage[figuresright]{rotating}
\usepackage{url}
\usepackage{color}
\usepackage{subfig}
\usepackage{verbatim}
\usepackage{amsmath}
\usepackage{eurosym}
\begin{document}

\begin{frontmatter}


%%\title{Title\tnoteref{label1}}

%% e.g. \dochead{17th International Conference on Dynamical Processes in Excited States of Solids}

\title{Science prospects for SPHiNX - a small satellite GRB polarimetry mission}
%The design and performance of 


% As usual, a discussion of the author list is needed.
\author[a,b]{M. Pearce\corref{cor1}}
\author[a,b]{L. Eliasson}
\author[a,b]{N. Kumar Iyer}
\author[a,b]{M. Kiss}
\author[a,b]{R. Kushwah}
\author[a,b]{J. Larsson}
\author[a,b]{C. Lundman}
\author[a,b]{\mbox{V. Mikhalev}}
\author[a,b]{F. Ryde}
\author[a]{T.-A. Stana}
\author[f]{H. Takahashi}
\author[a,b]{F. Xie}

%\author[a,b]{\corref{cor1}}
%\author[a,b]{}

\address[a]{KTH Royal Institute of Technology, Department of Physics, 106 91 Stockholm, Sweden}
\address[b]{The Oskar Klein Centre for Cosmoparticle Physics, AlbaNova University Centre, 106 91 Stockholm, Sweden}
\address[f]{Hiroshima University, Department of Physical Science, Hiroshima 739-8526, Japan}
\cortext[cor1]{Corresponding author. pearce@kth.se}

%%%%%%%%%%%%%%%%%%%%%%%%%%%%%%%%%%%%%%%%%%%%%%
\begin{abstract}

Gamma-ray bursts (GRBs) are exceptionally bright electromagnetic events occurring daily on the sky. 
The prompt emission is dominated by X-/$\gamma$-rays.
Since their discovery over 50~years ago, GRBs are primarily studied through spectral and temporal measurements.
The properties of the emission jets and underlying processes are not well understood.
A promising way forward is the development of missions capable of characterising the linear polarisation of the high-energy emission.
For this reason, the SPHiNX mission has been developed for a small-satellite platform. 
The polarisation properties of incident high-energy radiation (50--600~keV) are determined by reconstructing Compton scattering interactions in a segmented array of plastic and Gd$_3$Al$_2$Ga$_3$O$_{12}$(Ce) (GAGG(Ce)) scintillators. 
During a two-year mission, $\sim$200 GRBs will be observed, with $\sim$50 yielding measurements where the polarisation fraction is determined with a relative error $\leq$10\%. This is a significant improvement compared to contemporary missions.
This performance, combined with the ability to reconstruct GRB localisation and spectral properties, will allow discrimination between leading classes of emission models. 

\end{abstract}
%%%%%%%%%%%%%%%%%%%%%%%%%%%%%%%%%%%%%%%%%%%%%%

\end{frontmatter}
%
%\linenumbers
%
%%%%%%%%%%%%%%%%%%%%%%%%%%%%%%%%%%%%%%%%%%%%%%
\section{Introduction}

Gamma-ray bursts (GRBs) are the brightest electromagnetic events in the universe, occurring randomly on the sky approximately daily~\cite{paciesas1999,paciesas2012}. 
The emission is characterised by two epochs - the prompt phase, lasting for seconds to minutes and dominated by X-/$\gamma$-rays, and the afterglow, lasting for days and emitting at lower energies. The afterglow is relatively well understood~\cite{Zhang&Kumar}. Many aspects of the prompt phase remain unknown, but
the fireball model~\cite{Meszaros06} is a generally accepted scenario, where the GRB is formed during the collapse of a massive object into a black hole. 
Recent observations of gravitational waves~\cite{GW} in coincidence with a short GRB (typical prompt duration $<$2~s) have shown that the progenitor is a merger of two neutron stars. Neutron star-black hole mergers can also result in short GRBs.  For long GRBs (typical prompt duration up to 100~s), the progenitor is instead connected to broad line type 1c supernovae~\cite{Cano}.
Understanding the GRB prompt emission mechanism holds the key to using GRBs as probes of the early universe and of extreme physics such as relativistic jets, relativistic magneto-hydrodynamics, aberration of light, relativistic shock waves, and Lorentz invariance violation~\cite{context1,context2}.

During the process of collapse and merger, two highly relativistic jets of plasma are emitted along the rotational axis of the black hole.
A GRB is observed if one of the jets is directed to earth. 
Energy dissipation within the outflow (e.g. internal shocks or magnetic reconnection) gives rise to the prompt emission (keV--MeV). 
The prompt energy spectrum is featureless and is often well described by a smoothly broken power-law 
with a break at a peak energy, $E_p$, typically occurring in the range from a few tens to several hundred keV.
%in the range of 100--300~keV. 
The Band function is often used, smoothly connecting a low-energy power-law, with spectral index $\alpha$  
(the photon flux $N_E \propto E^\alpha$, where $E$ is the energy)
to a high-energy power-law with spectral index $\beta$~\cite{Band}. 
There is a large spread in the measured values of $\alpha$, $\beta$ and $E_p$ \cite{paciesas2012}, with a typical GRB having $\alpha=-1$, $\beta=-2.5$ and $E_p=$ 200~keV.
As the jet interacts with the surrounding medium it is decelerated producing a forward external shock, the emission from which forms the afterglow.

GRB detectors typically measure the energy distribution (spectra) and the arrival time (light-curve) of the GRB photons. Even though large samples of bursts have been observed, the properties of the jets and the underlying emission process remain poorly understood. 
The study of GRB jets is inherently difficult since images cannot be produced due to the large observing distance.
Measurements of the linear polarisation properties of the detected photons address this problem. Linear polarisation is described using two parameters: $(i)$ the polarisation fraction (PF, \%) describing the magnitude of beam polarisation; and, $(ii)$ the polarisation angle (PA, degrees) which defines the orientation of the electric field vector of the incident photon beam relative to, e.g., celestial north.

Despite the scientific value of GRB polarimetry, there is a lack of reliable observational data~\cite{mcconnell}. 
Between 2010 and 2012, the GAP polarimeter on-board the IKAROS spacecraft measured the polarisation of three bright GRBs in the 70-300~keV energy band~\cite{GAP1,GAP2}. The measurements indicated high PF values, and for the brightest burst, polarisation parameters were determined in two time bins revealing a 90$^\circ$ change in PA. The measurements had weak statistical significance and additional observations with more sensitive missions are required.
POLAR is a gamma-ray polarimeter mission (50--500~keV) launched with the Chinese Tiangong-2 space station in 2016~\cite{polar}. After $\sim$6~months of operations, data-taking ceased due to instrument malfunction. Polarisation data is expected for some of the $\sim$50 observed GRBs~\cite{Merlin}.
 The AstroSat mission was launched in 2015. The CZTI instrument is a general purpose coded aperture spectrometer for X-ray observations which can be used for polarimetry.
 During the first year of operations, 47 GRBs were detected and polarisation parameters determined for 11 of the brightest bursts~\cite{astrosatGRB}. 
High PF values were observed for the majority of GRBs, albeit with relatively large uncertainties, as discussed in Section~\ref{sec:comp}. 
Observations have also been reported from instruments not designed for polarimetry, e.g. the INTEGRAL~\cite{integralGRB1,integralGRB2} and RHESSI~\cite{rhessiGRB} missions. 
Since no polarimetric calibration was performed prior to observations, it is difficult to ascertain the reliability of the reported results~\cite{dispute1,dispute2,dispute3}.   

This paper describes the Satellite Polarimeter for High eNergy X-rays (SPHiNX) -- a satellite-borne instrument for hard X-ray polarimetric studies of GRBs. 
The instrument design is optimised for polarimetry and with a large field-of-view, $\sim$120$^\circ$ opening angle, and collecting area, $\sim$800~cm$^2$, a large sample of $\sim$200 GRBs will be provided during the two year mission. The light-curve and spectral shape will be determined for all GRBs.    
Polarisation parameters will be reconstructed in the energy range 50-600~keV for $\sim$50 GRBs. The arrival time of X-rays will be determined with an absolute timing accuracy of 1~ms to allow correlation with other missions. These instrument characteristics will allow discrimination between different classes of GRB emission models.

This paper is organised as follows. 
In Section~\ref{sec:motivation}, the scientific motivation for the mission is presented. 
An overview of the mission parameters and constraints is presented in Section~\ref{sec:mission}. 
The instrument design is described in Section~\ref{sec:design} and the operational and calibration strategy is outlined in Section~\ref{sec:ops+calib}. 
The simulated instrument characteristics and scientific performance are discussed in Section~\ref{sec:performance}.
An outlook is presented in Section~\ref{sec:outlook}.

%%%%%%%%%%%%%%%%%%%%%%%%%%%%%%%%%%%%%%%%%%%%%%
\section{Scientific motivation}
\label{sec:motivation}
The objectives of the SPHiNX mission are to identify the properties of GRB jets, and to identify the mechanism behind the high-energy emission. It is not known whether the emission is produced far down in the jet where the densities are high and the photons and plasma are closely connected (photospheric emission) or whether the emission is produced at large distances from the progenitor, where turbulence and shocks are responsible for the energy release (optically thin emission). Likewise, the magnetisation of the jet is a fully open question which depends on how the jet is formed and launched. Finally, it is unclear whether the jets are symmetric, if they are wide or narrow, or if they have a varying lateral profile. 
These jet characteristics are encoded into the energy spectrum and polarisation properties of the observed emission.
The resulting three measurement goals, summarised in Table~\ref{table:goals},  are described in the remainder of this section. 
These goals drive the design of the mission and the polarimeter, as outlined in Sections~\ref{sec:mission}--\ref{sec:ops+calib}.
Section~\ref{sec:performance} describes the ability of SPHiNX to discriminate between different classes of GRB prompt emission models.  
\begin{table}
\caption{SPHiNX measurement goals.}
\begin{center}
\begin{tabular}{|c|c|}
\hline
{\bf GRB property} & {\bf Measurement} \\
\hline
(1) Jet structure & \makecell{Time evolution of PA \\ within a burst}    \\
\hline
(2) Jet magnetisation & \makecell{Distribution of PF \\ for population of bursts}  \\
\hline
(3) Emission mechanism & \makecell{Distribution of PF \\ Energy dependence of PF \\ Time evolution of PF} \\
\hline
\end{tabular}
\end{center}
\label{table:goals}
\end{table}

\subsection{Geometric structure of GRB jets}
GRB jets are commonly modelled as axisymmetric structures, where the observed emission can only be polarised parallel or perpendicular to the sky projection of the jet axis~\cite{LundmanMNRAS}. A key prediction of the axisymmetric jet scenario is that PA will either change by 90$^\circ$ during the prompt GRB emission, or remain invariant. Other possibilities for the geometric structure include jets with internal structure, such as fragmented jets, or mini-jets within the larger jet~\cite{minijet}.
The mini-jet brightness is expected to change during the prompt emission, leading to pulsed emission from different parts of the jet. As the emission at a given time is dominated by the brightest mini-jet, PA is expected to fluctuate in a random fashion.

\subsection{Magnetisation of GRB jets}
Depending on the details of the jet launching mechanism, the plasma may be highly magnetised close to the central black hole. In this case, as the jets are launched, the magnetic fields are advected outwards, resulting in an ordered magnetic-field structure across the entire jet. Such a jet will produce
synchrotron emission as electrons gyrate in the magnetic fields. Due to relativistic aberration of light, the observer can only see a small patch of the jet emission region. Within this patch, the magnetic field is ordered, and, as a consequence, most observed GRBs will be highly polarised. The maximum PF is expected to be $\sim$50\%, while a typical value for most observers is $\sim$40\%~\cite{Luyitkov2003}.
Some jet launching mechanisms instead predict low magnetisation of the jet~\cite{magnetisation}. 
Shocks within the jet can still produce weak magnetic fields, but the direction of the field lines will vary randomly on small scales. This causes much of the observed jet patch polarisation signal to average out for on-axis observers. Observers that have the jet edge within their field-of-view can still see a substantial PF due to the asymmetric shape of the observed patch. In the weakly magnetised scenario, the distribution of PF will thus peak at 0\%, but a tail extends to high PF values due to viewing angle effects. 

\subsection{High-energy emission mechanism}

Different mechanisms can dominate the GRB prompt phase emission. 
Depending primarily on the dimensionless entropy and the magnetisation of the outflow, the emission can stem from the photosphere, internal shocks, magnetic reconnection, or the external shock~\cite{Z&M,B&B}. 
The high-energy emission is thus typically attributed to synchrotron radiation~\cite{R&M1994,Granot03} or emission from the photosphere~\cite{Goodman1986,Paczynski1986,Lundman2014}. 
The Compton drag model is also considered~\cite{Lazzati04}, but is not as well established. 
If the jets are highly magnetised, synchrotron emission is expected to dominate which will be revealed through a measurement of PF. 
If the jet has low magnetisation, discrimination 
is still possible by characterising the tail of the PF distribution.

For synchrotron models, softer emission is expected to be more polarised, resulting in a correlation between $\alpha$ and PF. Furthermore, the allowed values of $\alpha$ are restricted to $\alpha \leq -\frac{2}{3}$~\cite{Preece}. For photospheric emission the spectrum can be significantly harder, with $\alpha$ approaching unity. 
Two mechanisms can broaden the spectrum away from a black-body: energy dissipation below the photosphere and geometric effects~\cite{Lundman2013}. In both scenarios, polarised emission is expected for observers viewing the jet off-axis. The maximum predicted PF is $\sim$40\%. 
A correlation between $\alpha$ and PF is also predicted in photospheric models. 
For energy dissipation below the photosphere, only photons below the peak energy (synchrotron photons) will be polarised. 
Studying PF as a function of photon energy is therefore a useful discriminator. 
In the Compton drag model, the high-energy emission is produced due to up-scattering of the progenitor star photons by the jet. In this model, polarised emission is also produced as a viewing effect. The spectrum of the observed emission is determined by the range of temperatures in the progenitor star photon field, and is therefore not expected to correlate with the polarisation. The predictions are summarised in Table~\ref{table:correlate}.
\begin{table*}
\caption{Predictions for the polarisation fraction (PF) and spectral index ($\alpha$) for four emission models: (i) synchrotron emission in an ordered magnetic field (SO), (ii) synchrotron emission in a random magnetic field (SR), (iii) photospheric jet emission (PJ), and, (iv) Compton drag (CD).}
\begin{center}
\begin{tabular}{|l|c|c|c|c|}
\hline
 		& {\bf SO} & {\bf SR} & {\bf PJ} & {\bf CD} \\
\hline
Maximum PF 		& $\sim$50\% & $\sim$40\% & $\sim$40\% & $\sim$90\% \\
PF peak 			& $\sim$40\% & $\sim$0\% & $\sim$0\% & $\sim$0\%\\
Allowed $\alpha$ 	& $\leq$-2/3 & $\leq$-2/3 & $\leq$1 & $\leq$0 \\
PF-$\alpha$ correlation & Negative & Negative & Negative & None \\
\hline
\end{tabular}
\end{center}
\label{table:correlate}
\end{table*}

%
%%%%%%%%%%%%%%%%%%%%%%%%%%%%%%%%%%%%%%%%%%%%%%
\section{Mission overview}
\label{sec:mission}

The SPHiNX mission concept is developed for the Swedish InnoSat small satellite platform~\cite{innosat} produced by OHB Sweden and
{\AA}AC Microtec. The Swedish National Space Agency initiated the InnoSat programme to provide a recurring launch opportunity for scientific payloads proposed by Swedish researchers. 

The platform set-up for SPHiNX is detailed in Table~\ref{table:innosat}. 
The proposed orbital inclination, 53$^\circ$, is the lowest available and is partly motivated by the expected availability of piggy-back launch opportunities. 
Platform structural limits and the attitude control system dictate the maximum payload mass.
The attitude control system ensures that the SPHiNX field-of-view is oriented with quasi-zenith pointing while also maintaining a sun-pointing direction for the solar panel. 
Orbit analysis shows that the earth intrudes the polarimeter field-of-view $<$1\% of the year by a maximum of 7$^\circ$. 
\begin{table}
\caption{The InnoSat platform specifications.}
\begin{center}
\begin{tabular}{|l|c|}
\hline
& {\bf Specification} \\
\hline
Orbital inclination & 53$^\circ$ \\
Orbital altitude & 550 km \\
Launch type & Piggy-back\\
Nominal mission duration & 2 years \\ 
\hline
Maximum payload mass & 25 kg\\
Maximum payload volume & 480 $\times$ 525 $\times$ 700 mm$^3$ \\
Maximum payload power & 30 W\\
\hline
Daily downlink (one pass) & 150 MByte/day\\
\hline
Pointing & \makecell{Quasi-zenith \\ 3-axis stabilised \\ 0.1$^\circ$ precision}  \\
\hline
\end{tabular}
\end{center}
\label{table:innosat}
\end{table}
The platform provides (i) a GPS pulse-per-second signal enabling photon time-tagging to be synchronised to Universal Time, and, (ii) a signal to mark passage through the South Atlantic Anomaly (SAA) region. Science and housekeeping data transferred to the platform are downlinked each day. 
 
%%%%%%%%%%%%%%%%%%%%%%%%%%%%%%%%%%%%%%%%%%%%%%
\section{Instrument design}
\label{sec:design}

%%%%%%%%%%%%%%%%%%
\subsection{Formalism}

SPHiNX operates in the hard X-ray band, where Compton scattering interactions are used to determine the polarisation of incident photons~\cite{general}.
The Compton scattering differential cross-section is described by the Klein-Nishina relationship,
\begin{equation}
\frac{d\sigma}{d\Omega} =
\frac{1}{2}
r_{e}^{2}
\frac{k^2}{k_{0}^{2}}
\left(
\frac{k}{k_{0}} +
\frac{k_{0}}{k} -
2\sin^2 \theta\,\cos^2 \phi
\right),
\end{equation} 
where $r_e$ is the classical electron radius and $k_0$ and $k$ are the momenta of the incoming and scattered photon, respectively.
The polar scattering angle is given by $\theta$ and the azimuthal scattering angle, $\phi$, is defined relative to the plane of polarisation of the incident photon. 
X-rays will preferentially scatter in a direction perpendicular to the polarisation vector. 
The distribution of $\phi$ is a harmonic function referred to as a 'modulation curve', where the phase defines PA, and the modulation amplitude, $M$, divided by the value for a 100\% polarised beam, $M_{100}$, defines PF. 
Polarisation parameters can also be determined using unbinned Stokes parameters~\cite{Stokes} which avoids the introduction of systematic errors due to binning effects,
or Bayesian inference (particularly when there are few detected photons)~\cite{Mikhalev18}.

Counting rate asymmetries within the pixelated detector volume are used to determine the polarisation properties of GRBs.
Non-uniformities in the instrument response can result in a spurious polarisation signal.
Polarimeters designed for the observation of persistent sources can be  rotated around the viewing axis to mitigate this effect. 
The short observing times, along with the relatively large mean angle of incidence ($\sim$40$^\circ$, derived from a uniform distribution of GRBs on the sky),
make this approach impractical. 
Instead, the response to an unpolarised beam must be determined and the measured polarisation corrected for the residual modulation which results from the detector geometry and response variations~\cite{deconvolve}. This is achieved using computer simulations validated using both polarised and unpolarised beams in the laboratory~\cite{validate}. 
A particular challenge for large field-of-view polarimeters, such as SPHiNX, is that the polarimetric response depends on the incidence angle and energy spectrum of the GRB~\cite{muleri}. 
SPHiNX is designed to allow these parameters to be reconstructed.
When possible, these parameters can also be derived independently by other instruments which view the same GRB as SPHiNX. 
%%%%%%%%%%%%%%%%%%
\subsection{Overview of design}

\begin{figure}[!ht]
\centering
\includegraphics[width=8 cm]{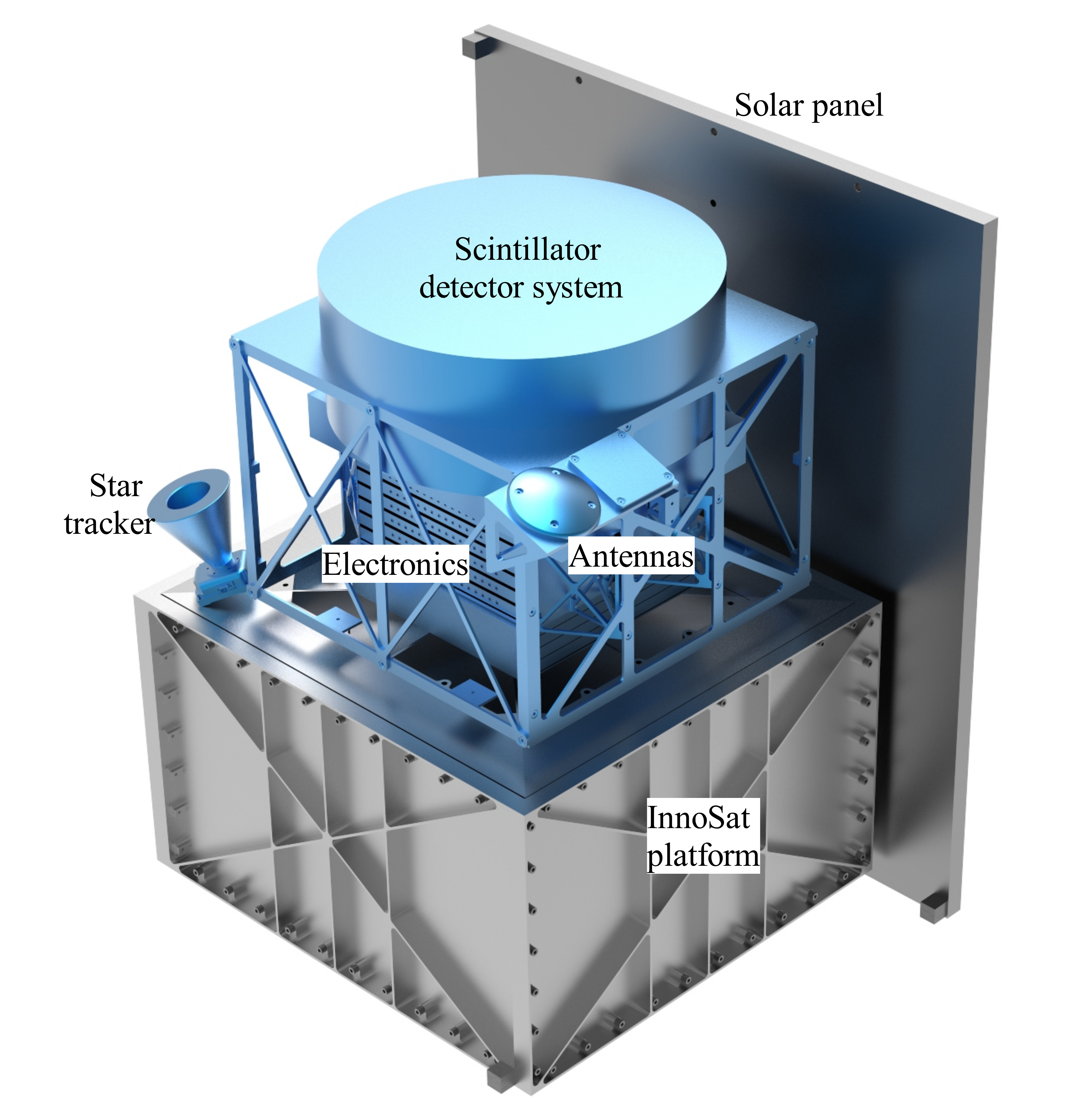}
\caption{The SPHiNX hard X-ray polarimeter installed onto the InnoSat platform. The InnoSat platform has dimensions 58$\times$53$\times$34~cm$^3$. The solar panel is 83~cm tall. Figure courtesy of OHB Sweden.}
\label{fig:overview}
\end{figure}

The SPHiNX polarimeter is shown mounted on the InnoSat platform in Fig.~\ref{fig:overview}.
The polarimeter comprises a scintillator-based detector system connected to a front-end electronics system, a data acquisition system and power supply units. The various subsystems are mounted in a mechanical support structure.
Each of these components is described in the following sections. 
%
%%%%%%%%%%%%%%%%%%
\subsection{Scintillator detector system}

\begin{figure}[!ht]
\centering
\includegraphics[width=8 cm]{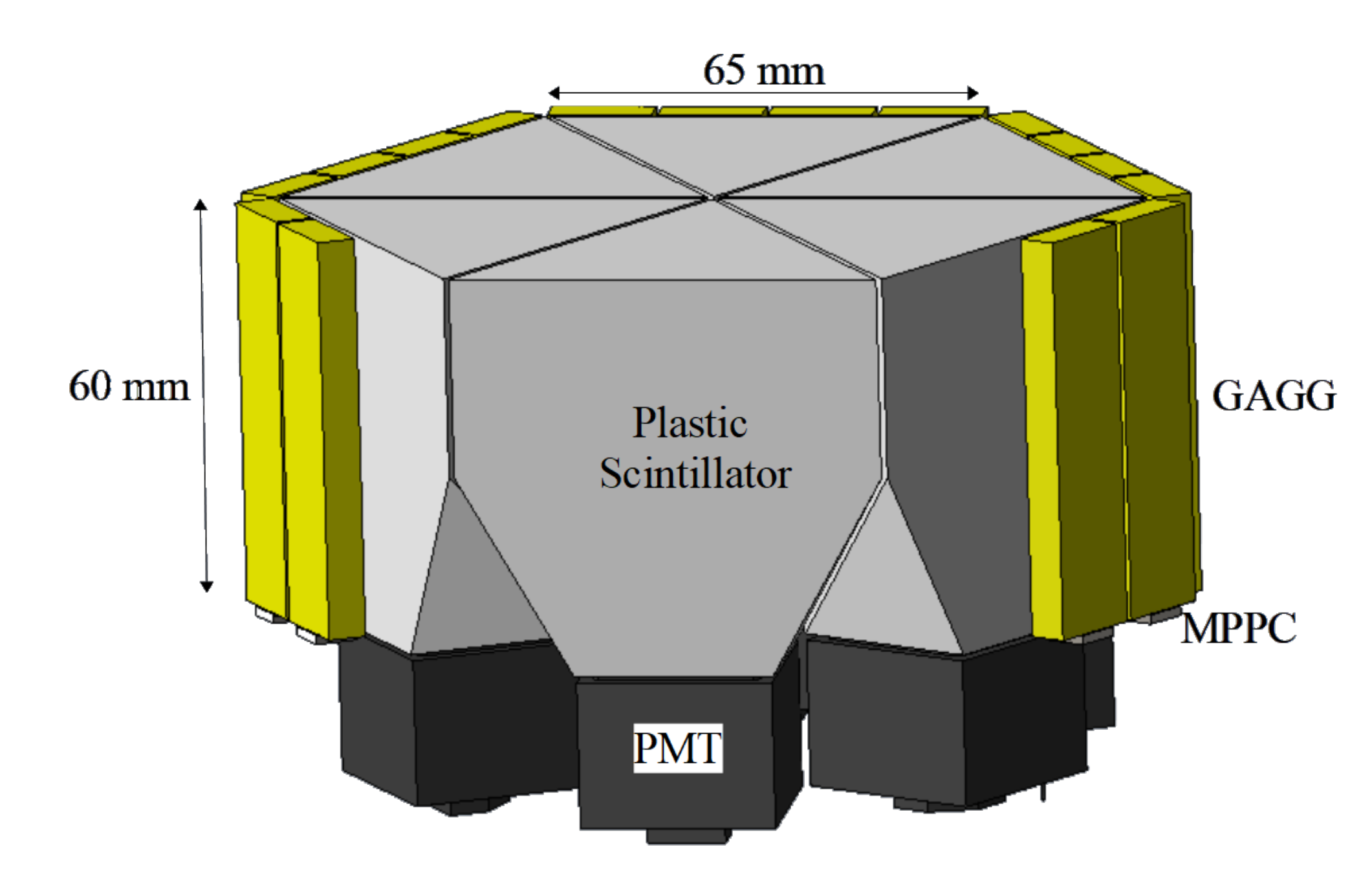}
\caption{The geometry of the plastic and GAGG scintillators. The plastic scintillator piece transitions from a triangular shape (one sixth of a hexagon) to the square cross-section of the PMT.}
\label{fig:scintgeom}
\end{figure}

The detector volume comprises an array of $(i)$ plastic scintillators, which give a high probability for Compton scattering (low atomic number, $Z$); and, $(ii)$ GAGG (Gadolinium Aluminium Gallium Garnet (Cerium doped) - Gd$_3$Al$_2$Ga$_3$O$_{12}$(Ce)) scintillators, which provide a high probability of photoelectric absorption (high $Z$). Scintillator properties are summarised in Table~\ref{table:scintillators}, and the scintillator geometry is shown in Fig.~\ref{fig:scintgeom}.
\begin{table*}
\centering
\caption{Scintillator properties.}
\label{table:scintillators}
\begin{tabular}{|l|c|c|}
\hline
					& {\bf Plastic scintillator} 	& {\bf GAGG}   \\
\hline
Supplier/type 			&	Eljen Technology EJ-204& C\&A        \\ 
Light-yield ($\gamma$/MeVee) & 1$\times$10$^{4}$  &    5.6$\times$10$^{4}$              \\ 	
Decay-time (ns)		 	& 	1.8	&     88             \\  
Density (g/cm$^{3}$) 	&	1.02				& 6.63                  \\ 
Peak wavelength (nm) 	&	408				&       520           \\ 
Refractive index & 1.58 & 1.9 \\	
\hline
\end{tabular}
\end{table*}

Plastic scintillator is a well established detector material which has been used in many space missions, and is not discussed further here. Conversely, GAGG is a relatively new development. It has high density with correspondingly high stopping power, high light-yield (e.g. comparable to CsI(Tl) and exceeding BGO by a factor of five), fast scintillation decay-time (88~ns), and is non-hygroscopic. 
The temperature dependence of the scintillation light-yield and decay-time is small, $\sim$10\%~\cite{GAGGSpace}. 
The radiation tolerance of detector components must be determined for space missions. 
Light-yield degradation of $\sim$10\% has been reported for GAGG after a 100~krad ionising dose at a $^{60}$Co facility ($>$10~years in low-earth orbit)~\cite{GAGGSpace}. 
A study of the activation characteristics for low-earth orbit has shown that GAGG has the lowest activation-induced count rate, for the energy range 30--400~keV, when compared to other commonly used inorganic scintillators~\cite{GAGGcalib}. 
A number of undesirable properties have also been identified.
The reported presence~\cite{GAGGSpace} of a prominent annihilation line at 511~keV may dictate the upper energy limit for polarimetry and therefore warrants further study. 
The light-yield is reported to become non-linear below 30~keV~\cite{GAGGSpace}, requiring a careful laboratory-based calibration campaign prior to launch. 
Large energy deposits from cosmic-ray electrons (e.g. in the SAA) will generate large amounts of scintillation light. 
This may lead to afterglow which will deteriorate the energy resolution. 
Compared to other inorganic scintillators, the GAGG afterglow has a relatively high intensity and a long decay-time. 
As a part of manufacturer acceptance procedures, GAGG elements can be irradiated with UV light to identify samples which are more susceptible to induced afterglow, e.g. due to crystal lattice imperfections arising during the production process.

\begin{figure}[!ht]
\centering
\includegraphics[width=8 cm]{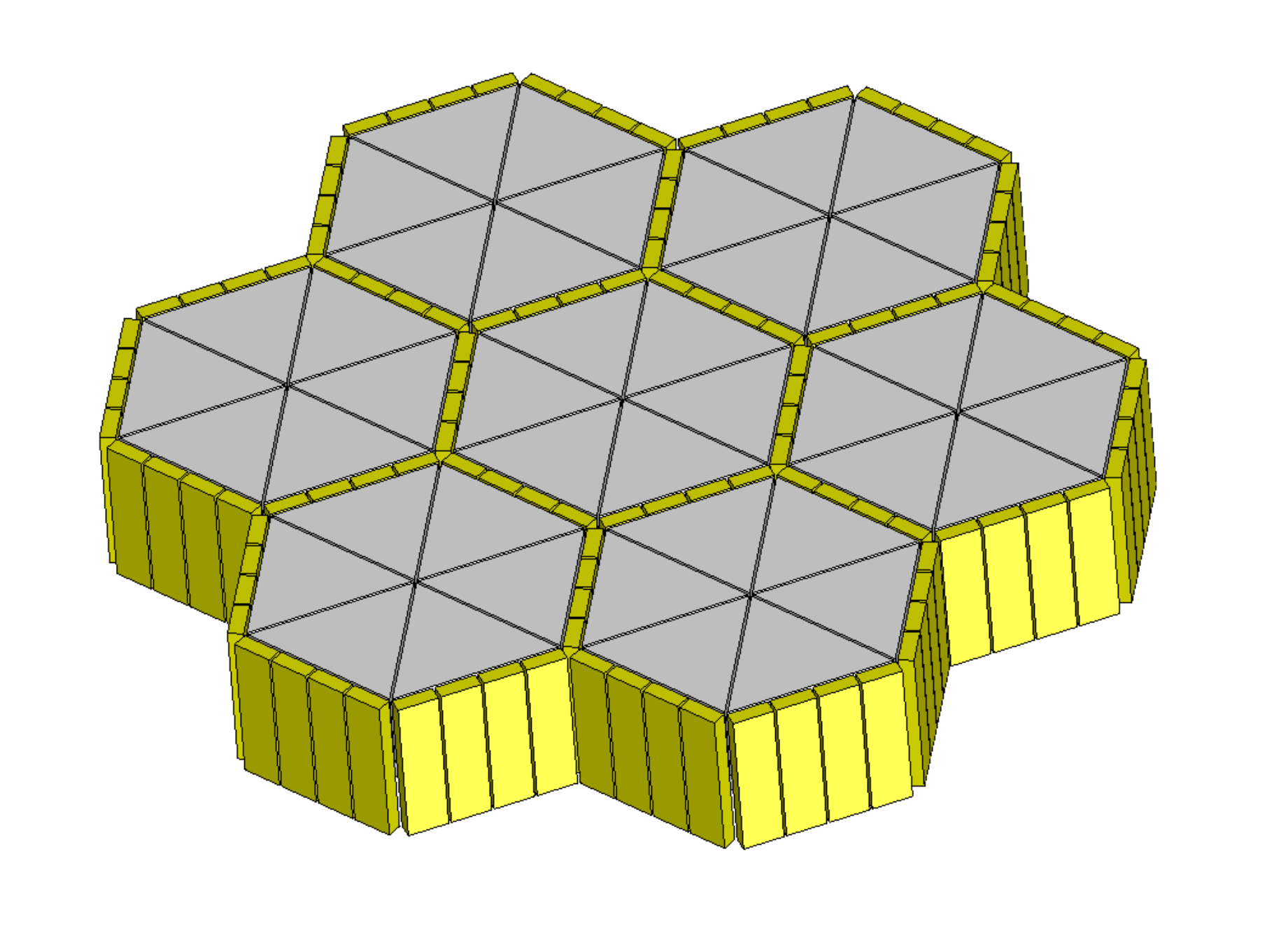}
\caption{The scintillator detector array. Plastic scintillators are shown in grey and GAGG scintillators in yellow. The array has a width of $\sim$40~cm at the widest point.}
\label{fig:array}
\end{figure}

The detector array, illustrated in Fig.~\ref{fig:array}, is $\sim$40~cm wide, with a geometric area of $\sim$800~cm$^2$. It consists of 42 plastic scintillators forming seven hexagons, each comprising six pieces. Each plastic hexagon is surrounded by a sub-divided wall of GAGG scintillators. In order to reduce mass and the number of read-out channels, adjacent plastic hexagons share a common wall, resulting in a total of 120 GAGG scintillators. The splitting of hexagons into separate scintillators and the GAGG segmentation allows reconstruction of the azimuthal scattering angle and provides redundancy against component failure.
The azimuthal scattering angle of incident X-rays is defined by two-hit events (two coincident energy deposits in separate scintillators), primarily a Compton scatter in a plastic scintillator, followed by photoelectric absorption in a GAGG scintillator. 
The inclusion of plastic-plastic events and GAGG-GAGG events is beneficial since event statistics are improved and the polarimetric sensitivity increases at the lower and upper extremes of the SPHiNX energy range, respectively. 

In the baseline design, each of the six plastic elements which form a hexagonal scatterer is individually read out by a small form factor (30~mm $\times$ 30~mm $\times$ 22~mm; 33~g) single anode Hamamatsu R7600U-200 photomultiplier (PMT). This arrangement eliminates the optical cross-talk between scintillator units which would arise if a large area multi-anode PMT was used to read out the entire hexagonal scintillator assembly.
Read-out based on multi-pixel photon counter (MPPC) arrays is also under consideration.

The MPPC comprises an array of avalanche photodiode pixels, operating in Geiger-mode above the breakdown voltage. 
The output signal from the MPPC is the analogue sum of the triggered pixels and is thereby proportional to the number of detected photons. 
MPPCs have a quantum efficiency comparable to a PMT. The MPPC offers several advantages over a PMT, e.g. $(i)$ it is a compact and robust solid-state device; $(ii)$ the operating voltage required to obtain a representative gain of 10$^6$ is significantly lower ($\sim$50~V rather than $\sim$900~V); $(iii)$ the device characteristics are not affected by magnetic fields. A particular challenge is that the dark counting rate is typically higher than a PMT. This can be mitigated by using a short coincidence window ($\sim$100~ns) when identifying two-hit events. The temperature dependence of the gain requires regulation of the bias voltage in-orbit based on data provided by temperature sensors.    
There is little experience of using MPPCs in the space environment. One concern is radiation tolerance. For the low-earth orbit foreseen, the effect of ionising radiation is negligible~\cite{G400}. Bulk displacement damage due to protons and neutrons may cause the dark current to increase by up to an order of magnitude, but there is only a small decrease in the pulse amplitude~\cite{Bloser}. For a bright scintillator like GAGG, this is not expected to be problematic since selection thresholds applied to pulses can be relatively high. 

A monolithic MPPC, Hamamatsu S13360-6050PE, is foreseen for the read-out of the GAGG elements. 
The small form factor, 7.4~mm $\times$ 6.9~mm $\times$ 1.5~mm, and low mass, 1~g, permit the required close-packed geometry. 

The periphery of the scintillator array is surrounded by a multi-layer metal shield, primarily to mitigate the cosmic X-ray background and low energy charged particles. The shield comprises 1~mm lead (outer-most layer), 0.5~mm tin, and 0.25~mm copper (inner-most layer), for an average density of 7.29 g/cm$^3$. 
The aperture of the scintillator array and the shield are covered by a 1~mm thick carbon-fibre reinforced plastic (CFRP) cap, which protects the scintillators and provides additional background rejection - especially for low energy charged particles. The earth albedo background is shielded by the InnoSat platform. 

%%%%%%%%%%%%%%%%%%
\subsection{Electronics}
\label{sec:electronics}

\begin{figure*}[!ht]
\centering
\includegraphics[width=12cm]{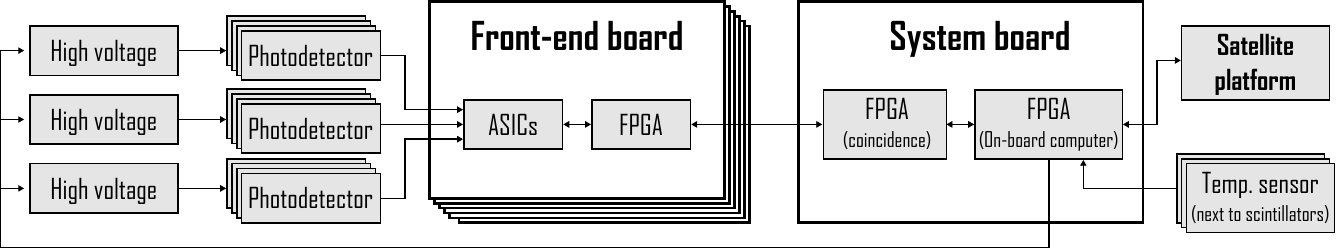}
\caption{Block diagram of the SPHiNX electronics.}
\label{fig:elec-block-diag}
\end{figure*}

A schematic overview of the SPHiNX electronics system is presented in Fig.~\ref{fig:elec-block-diag}. 
Output signals from PMTs and MPPCs are processed by application-specific integrated circuits (ASICs) housed on Front-End Boards (FEBs). 
A field-programmable gate array (FPGA) on each FEB provides configuration and control signals for the ASICs and generate signals required for event selection (see Section~\ref{sec:selection}).
The FEBs are connected to one or more system boards (SBs) housing the coincidence and selection logic required to identify two-hit events.
The SB also houses the On-Board Computer (OBC) and provides the interface to the InnoSat platform.

Power consumption requirements and space constraints motivate the adoption of an ASIC-based system.  
Several ASICs are currently under evaluation, the IDE3380 'SIPHRA' from IDEAS (Norway) and the 'Citiroc' and 'Catiroc' from Weeroc (France). 
The ASICs operate in a similar fashion, providing pre-amplification, shaping and track-and-hold functions. Resulting pulses are digitised using discrete analog-to-digital converters.    

The photo-sensors require a bias voltage ($\sim$900~V for PMTs and $\sim$50~V for MPPCs).
The power supply systems will be controlled by the OBC allowing the photo-sensor bias to be regulated to achieve gain uniformity across the detector array. 
This is particularly important since the MPPC gain and GAGG light-yield are temperature dependent.
The temperature in the vicinity of the scintillator detectors is monitored and the bias voltage is changed accordingly during the orbit. 
Since the polarimeter is thermally isolated from the platform and is located behind the sun-pointing solar panel, a relatively stable thermal environment is achieved. Simulations indicate that temperature fluctuations of 1$^\circ$C are expected during an orbit. The detector array is thermally decoupled from the electronics enclosures to ensure a nominal operating temperature not exceeding room temperature.  

%%%%%%%%%%%%%%%%%%
%
\subsection{Power and mass budgets}
For the baseline design of SPHiNX, the power consumption of the electronics is estimated to be 26.8~W, including a 20\% margin, as described in Table~\ref{table:power}. 
\begin{table}
\caption{SPHiNX power budget.}
\begin{center}
\begin{tabular}{|l|c|}
\hline
{\bf Component}	& {\bf Power (W)} \\
\hline
12 $\times$ FEB	& 	10.1	\\
System Boards		& 	3.1	\\
Power supply		& 	4.1	\\
\hline
Subtotal			&	22.3	\\
\hline
20\% contingency	& 	4.5	\\
\hline
Total			&	26.8  \\
\hline
\end{tabular}
\end{center}
\label{table:power}
\end{table}
The overall mass, assuming the mechanical design shown in Fig.~\ref{fig:overview}, is estimated to 23.2~kg, as detailed in Table~\ref{table:mass}. 
%The estimated uncertainty on this value is $\pm$1.7~kg.
In both cases, the requirements detailed in Table~\ref{table:innosat} are satisfied.
\begin{table}
\caption{SPHiNX mass budget.}
\begin{center}
\begin{tabular}{|l|c|}
\hline
{\bf Component}	& {\bf Mass (kg)} \\
\hline
Polarimeter	& 	9.6	\\
Shielding		& 	1.8	\\
Structure		& 	4.5	\\
Electronics	& 	7.3	\\
\hline
Total			&	23.2 \\
\hline
\end{tabular}
\end{center}
\label{table:mass}
\end{table}
%
%%%%%%%%%%%%%%%%%%%%%%%%%%%%%%%%%%%%%
\section{Operations and calibration}
\label{sec:ops+calib}

SPHiNX has a number of operation modes: (i) initialisation, (ii) operations mode (data acquisition or calibration), (iii) SAA mode, (iv) low power mode, and, (v) safe mode.
After being powered on by the platform, the initialisation procedure configures electronic registers and photo-sensor bias voltages. 
Once configured, the polarimeter can be placed in operations mode for either data acquisition or calibration operations. 
Continuous or triggered data acquisition modes are foreseen. In both cases, 
multiple- ($\geq$ 2) hit events are recorded, along with a fraction of one-hit events. Pedestal levels for each read-out channel are recorded at 1~Hz rate. In the continuous mode, data is always recorded allowing background conditions to be assessed. In the triggered mode, GRBs are identified through the instantaneous increase in the detector counting rate. A ring buffer allows data to be collected immediately prior to the GRB for background level estimation.
For calibration, one-hit pulse height spectra will be acquired for all scintillator channels in order to determine the detector energy scale. Calibration data may be derived from weak radioactive sources embedded in the scintillator array. Another possibility is to use minimum ionising particles, activation lines of GAGG ($\sim$50~keV and $\sim$150~keV)~\cite{GAGGcalib}, or 511~keV annihilation radiation from cosmic-ray e$^{+}$ interactions. 
The high particle fluxes present in the SAA may cause electrical discharges in the PMTs. An SAA mode is therefore implemented where-by the PMT high voltage is reduced while a 'crossing SAA' signal is asserted by the platform. 
If safe mode is requested by the platform, the polarimeter is prepared for power off, e.g. all photo-sensor voltages are reduced to zero, all computer/memory operations are terminated, etc. Safe mode heaters for component survival are controlled by the platform. Housekeeping data are created periodically in all modes except safe mode.

%%%%%%%%%%%%%%%%%%%%%%%%%%%%%%%%%%%%%%%%%%%%%%
\section{Performance studies}
\label{sec:performance}
%
%%%%%%%%%%%%%%%%%%
\subsection{Simulation set-up}
\label{sec:sim_intro}

The performance of the SPHiNX design has been evaluated using Geant4 (version 4.10.02.p02)~\cite{G4}. 
The \texttt{Shielding} physics list is used which includes electromagnetic physics, hadronic physics and decay physics (radio-activation of materials). 
For low-energy electromagnetic processes, including polarisation, the
\texttt{G4EmLivermorePolarizedPhysics} physics list is also included.
The geometry of the Geant4 mass model is derived from the SPHiNX CAD model (Fig.~\ref{fig:overview}), with simplifications to reduce computation time but with sufficient accuracy to allow the background interactions to be studied in detail. All significant components are implemented, including the solar panel, scintillator detectors, photo-sensors, electronics boards, shielding materials, mechanical support structure, and the InnoSat platform mechanics, including the main internal components.
The simulation considers ideal photo-sensors and read-out electronics.
Effects such as dark currents, electronics read-out noise, attentuation of scintillation light, etc. are not implemented.
These effects may degrade the spectroscopic performance of SPHiNX and must be fully characterised during pre-launch calibration studies. 
The effect on polarisation measurements can be minimised through a judicious choice of energy thresholds when finalising the selection of two-hit events. 

%%%%%%%%%%%%%%%%%%
\subsubsection{Event selection}
\label{sec:selection}

In order to reduce background, selection criteria are applied to scintillator energy deposits to define events which are used for spectroscopy, timing and localisation (one-hit events) or polarimetry (two-hit events). 
The following criteria must be fulfilled:
\begin{itemize}
\item Each energy deposit must exceed the {\it hit threshold}.
\item At least one hit has an energy deposit above the {\it trigger threshold}.
\item No hit has an energy deposit exceeding the {\it upper threshold}.
\end{itemize}
Noise arising from the photo-sensors and read-out electronics is rejected by the hit threshold.
The 5~keV threshold corresponds to the energy deposited by an incident 50~keV photon which Compton scatters through an azimuthal angle of 90$^\circ$. 
The resulting low energy deposit in plastic drives the required sensitivity of the read-out electronics.
The trigger threshold is defined to be less than the energy that a 50 keV incident photon retains after Compton scattering through an angle of 180$^\circ$.
The upper threshold allows minimum ionising particles passing through the scintillator array to be rejected, but retains 511~keV energy deposits from 
e$^+$ annihilation which may be used for energy scale calibration.
The thresholds implemented for the plastic and GAGG scintillators are detailed in Table~\ref{table:thresholds}. 
\begin{table}
\caption{Summary of thresholds.}
\begin{center}
\begin{tabular}{|l|c|c|}
\hline
			& Plastic (keV) & GAGG (keV) \\
\hline
Hit			& 	5	&	30	\\
Trigger		&	25	& 	30	\\
Upper 		& 	600	&	600	\\
\hline
\end{tabular}
\end{center}
\label{table:thresholds}
\end{table}

%%%%%%%%%%%%%%%%%%
\subsection{Polarimeter characteristics}
\label{sec:char}

\begin{figure}[!h]
	\centering
		\includegraphics{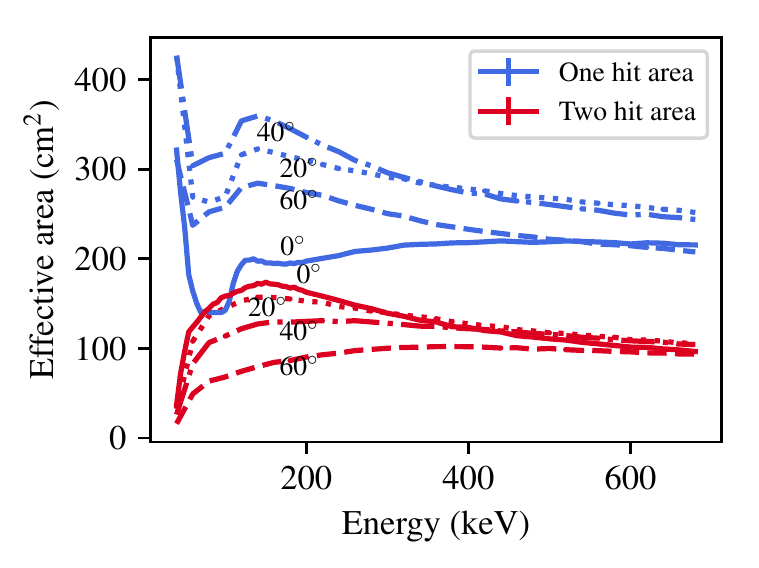}
	\caption{The energy dependence of effective area for one-hit
	(blue) and two-hit (red) events for a range of incidence angles.}
	\label{fig:effarea}
\end{figure}

\begin{figure}[!h]
	\centering
		\includegraphics{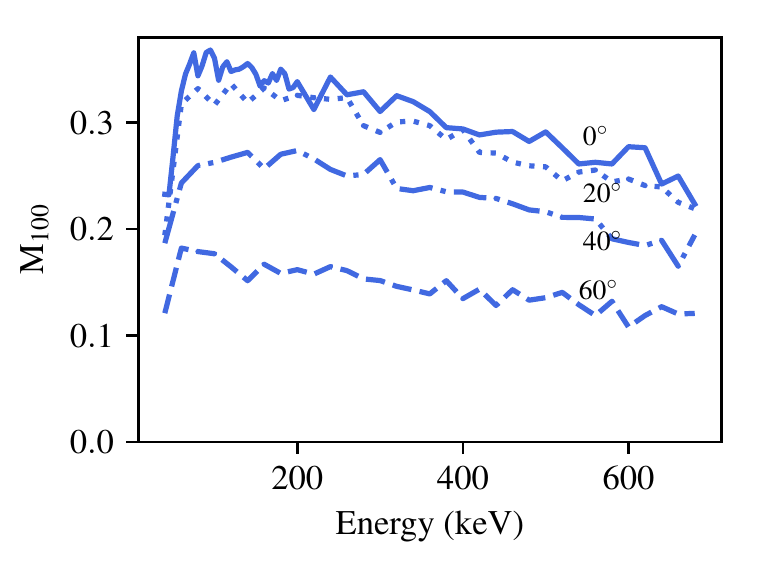}
	\caption{The energy dependence of $M_{100}$ for a range of incidence angles.}
	\label{fig:m100}
\end{figure}

A beam of mono-energetic photons which illuminates the entire scintillator array is used to determine the energy dependence of the effective area, shown in Fig.~\ref{fig:effarea}. 
The effective area is computed for both one-hit and two-hit events for a range of incidence angles. 
The energy dependence of the one-hit effective area demonstrates that plastic scintillator interactions dominate for energies lower than $\sim$50~keV, while GAGG interactions dominate at higher energies. 
The transition between these two regimes creates a dip in the one-hit effective area around 50~keV. 
The one-hit effective area increases with incidence angle up to $\sim$40$^\circ$, since the GAGG surface area/depth seen by the incoming beam of photons grows. 
The two-hit effective area decreases with both incidence angle and energy (above $\sim$50 keV). 
Across the SPHiNX energy range, the effective area exceeds 100~cm$^2$ for normally incident beams, with a value of $\sim$125 cm$^2$ at the mean incident angle, $\sim$40$^\circ$, and mean energy, $\sim$145 keV. The former value is derived from simulated GRBs uniformly distributed on the sky. The latter value assumes a typical Band spectrum for incident GRBs.
The energy and incidence angle dependence of $M_{100}$ is shown in Fig.~\ref{fig:m100}. 
The $M_{100}$ drops by half near the edge of the field-of-view and by $\sim$30\% (relative) at the mean angle of incidence (40$^{\circ}$).
%

%%%%%%%%%%%%%%%%%%
\subsection{Polarimetric performance}
\label{sec:sim}

The Minimum Detectable Polarisation~\cite{MDP}, MDP, quantifies the sensitivity of a polarimeter.
The MDP is defined at 99\% ($\sim$3$\sigma$) confidence level as  
\begin{equation}
\mathrm{MDP} = \frac{4.29}{M_{100}\,R_s} \sqrt{\frac{R_s + R_b}{T}},
\label{eqn:mdp}
\end{equation} 
where $R_s$ ($R_b$) is the signal (background) rate (Hz) and $T$ is the duration of the observation (s).
Unpolarised radiation has a 1\% probability of generating PF $>$ MDP through statistical fluctuations. 
Measured PF values are positive definite and follow a Rice distribution. 
Therefore, unless the reconstructed PF $\gg$ MDP, polarisation parameters must be corrected for bias~\cite{Mikhalev18}. 

%%%%%%%%%%%%%%%%%%
\subsubsection{Background response}
\label{sec:bkgnd}
A detailed description of background studies for the SPHiNX mission is presented elsewhere~\cite{Fei} - a summary follows in this section. 
The prompt background outside the SAA comprises primary (protons, alpha particles, electrons and positrons) and secondary (protons, electrons, positrons) cosmic rays. Primary cosmic rays are accelerated by celestial sources and travel through the galaxy before reaching the earth. When the primary particles impinge on the atmosphere and interact with residual gas molecules, secondary cosmic rays are produced, including albedo components (gamma-/X-rays and neutrons). The isotropic cosmic X-ray background (CXB) is also considered. Energy spectra for these components are adapted from a previous study of background conditions for the HXMT mission~\cite{HXMT} which operates at the same altitude as SPHiNX but a lower inclination (43$^\circ$). 

Within the SAA, the low-energy trapped electron and proton rates are too high to allow GRB detection and the PMT bias voltage will be reduced to protect against electrical discharge. The time spent inside the SAA 
is estimated using SPENVIS~\cite{spenvis} as \mbox{$\sim$5.5~hours/day}, resulting in a duty cycle for GRB observations of $\sim$80\%. 
Passages through the SAA generate a delayed background, due to the decay of radioactive isotopes which are produced by trapped protons in the energy range from 100 MeV to 400 MeV~\cite{HXMT}. While the timescale for prompt background is $<1\mu$s, for the delayed background this can extend to hundreds of days for some isotopes.

The resulting prompt background levels, after applying event selections, are shown in Table~\ref{table:bkgrate}. 
The two-hit background rate is found to be dominated by the CXB and albedo gamma-/X-ray components.
The delayed background increases rapidly during the first month in orbit before saturating after 1~year to $\sim$190~Hz for two-hit events. Aluminium materials in the platform structure are the dominant source of the delayed background. 
The polarimeter shielding can be further optimised during payload integration studies to significantly reduce this background source. 
\begin{table*}[ht]
\caption{Prompt background rates.}
\begin{center}
\begin{tabular}{| c | c | c | c |}
\hline
{\bf Component} & {\bf One-hit rate (Hz)} & {\bf Two-hit rate (Hz)}  & {\bf Higher multiplicity (Hz)}\\
\hline
Cosmic X-ray		& 	1270 	&	195		&	38	\\
Albedo gamma		& 	398		&	113		&	31	\\
Albedo neutron		& 	14		&	5		&	3	\\
Primary particles	& 	16		&	5		&	3	\\
Secondary particles	& 	9		&	5		&	3	\\
\hline
Total				&	1707		&	323		&	78	\\
\hline
\end{tabular}
\end{center}
\label{table:bkgrate}
\end{table*}
%

%%%%%%%%%%%%%%%%%%
\subsubsection{GRB detection}
\label{sec:GRB}

The GRB detection performance is studied using a sample of 1967 GRBs, where the energy spectra, fluxes (derived from the time-averaged Band
function) and temporal properties are taken from the
Fermi-GBM catalogue~\cite{grbcat} up to 24th November 2017. 
For each GRB, photons are generated both unpolarised and with 100\% PF. 
The position of the GRB within the field-of-view and the PA value are randomised.
Two-hit events which satisfy the event selection criteria form a modulation curve.
The modulation curve derived from the unpolarised photon flux is used to correct that obtained for a polarised flux. 
When performing the same procedure for flight data, the GRB location, energy spectrum and light-curve may be derived directly from SPHiNX using the one-hit counting rate data or be derived from simultaneous measurements made by other missions.   

\begin{figure}[!ht]
	\centering
	\includegraphics{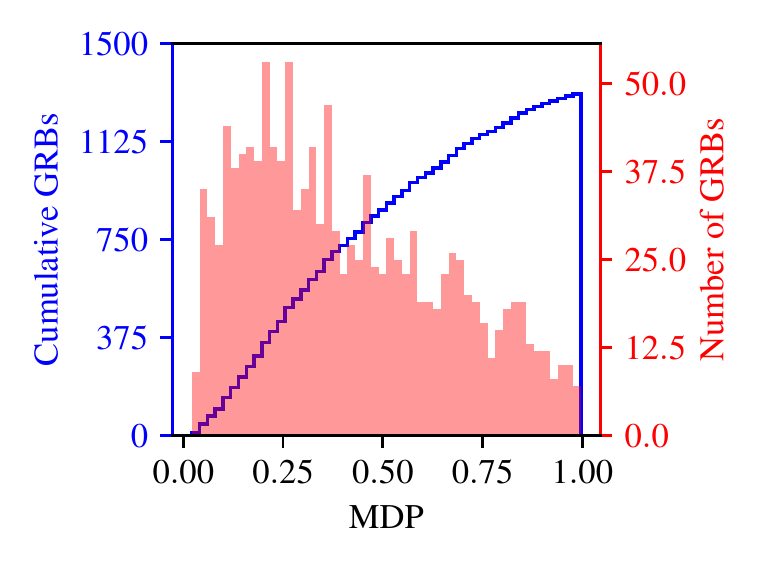}
	\caption{MDP distribution for 1967 GRBs in the Fermi-GBM catalogue. GRBs yielding MDP $>$ 1 are not shown.}
	\label{fig:mdpdistr}
\end{figure}

The $M_{100}$ obtained from the modulation curve is combined with the GRB flux, duration and background rate to define an MDP value. 
The resulting MDP distribution is shown in Fig.~\ref{fig:mdpdistr}. 
The background in flight is determined using data collected immediately before and after the GRB. 
For the studies reported here the background is estimated as detailed in Section~\ref{sec:bkgnd}.

\begin{figure}[!ht]
	\centering
	\includegraphics{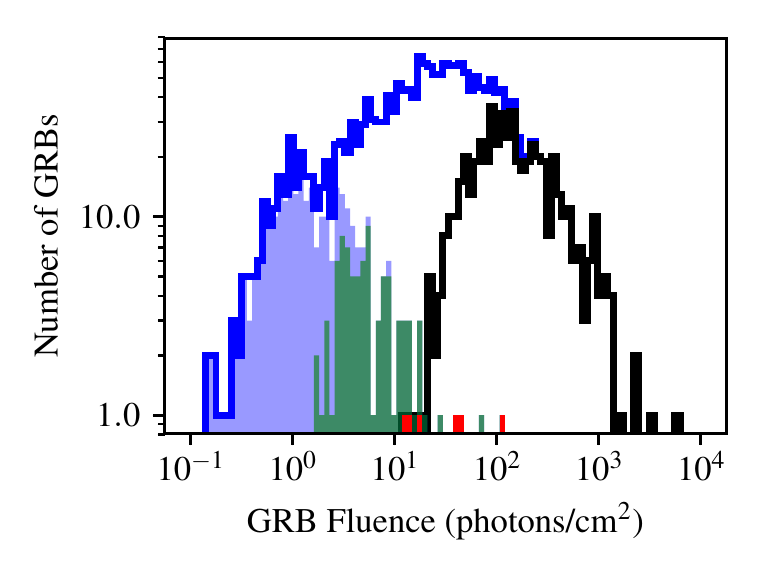}
	\caption{GRB fluence distribution from the
	entire Fermi-GBM catalogue (blue line) and for GRBs where the
	reconstructed SPHiNX MDP $<$ 0.3 (black line). The distribution for all short GRBs 
	(blue filled region), short GRBs with  MDP $<$ 1 (green filled region) and
	short GRBs  with  MDP $<$ 0.3 (red bars) are also shown.}
	\label{fig:n2h}
\end{figure}

Figure~\ref{fig:n2h} shows the fraction of GRBs from the Fermi-GBM catalogue for which SPHiNX
can make a statistically significant polarisation measurement, defined as MDP$<$0.3 (black curve).
The 1$\sigma$ error on PF can be approximated as MDP/3, i.e. $\leq$10\%.
A total of 532 GRBs satisfy this criterion, with a corresponding GRB fluence exceeding $\sim$50 photons/cm$^2$.  
Short GRBs account for only 1\% of the statistically significant GRBs, and the corresponding fluence limit is $\sim$10 photons/cm$^2$ (blue filled region).  As shown in the figure, while a large number of short GRBs yield MDP $<$ 1 (green filled region), very few yield MDP $<$ 0.3 (red bars).
%
%%%%%%%%%%%%
%
\subsection{Emission mechanism} 
\label{sss:sciper}
The scientific potential of SPHiNX is explored by determining the response to PF distributions expected from different emission models. 
The PF distributions for the synchrotron (SO and SR) and Compton Drag (CD) models are constructed through Monte Carlo sampling of Equations 11, 16 and 17 in \cite{toma2009} (hereafter TO09).
Parameter values follow Fig. 2 of TO09. 
The bulk Lorentz factor, $\Gamma$, is set to 100. 
The jet angle ($\theta_j$) distribution is described by Equation~18 of TO09. 
The cosine of the viewing angle, $\theta_v$, is uniformly distributed within the range $\cos(\theta_j + 5/\Gamma) < \cos(\theta_v) < 1.$

The PF distribution from the photospheric emission model (PJ) peaks at PF = 0 with small tails to higher PF values~\cite{Lundman2014}. 
The PF distribution is estimated assuming a geometric jet structure and considering the distribution of viewing angles and the variation of PF with $\theta_v$. 
Polarisation can be seen from these jets when viewing the edge of the jet in the relativistically beamed cone, i.e. 
for $\theta_j - 1/\Gamma \lesssim \theta_v \lesssim \theta_j + 1/\Gamma$~\cite{Lundman2014}.
The maximum PF occurs when $\theta_v = \theta_j$ for PF$_{\mathrm{max}}$ = 40\% and PF falls to zero for $\theta_v \gg \theta_j \pm 1/\Gamma$ \cite{Lundman2014}. 
This behaviour is approximated by a Gaussian distribution with a width, $\sigma_{PJ}$, of $1/\Gamma$, i.e. 
\begin{equation}
\mathrm{PF} = \mathrm{PF_{max}} \exp{\left( \frac{(\theta_v - \theta_j)^2}{2 \sigma_{PJ}^2} \right)}.
\end{equation}

The flux limit obtained from the Fermi-GBM catalogue (Fig.~\ref{fig:n2h}) is used to determine if SPHiNX will detect a given GRB flux, $F_G$. 
The flux for the Monte Carlo sampled GRBs is determined assuming a constant on-axis emission luminosity of $L_o = 10^{52}$~erg/s,
a luminosity distance $D_L$ obtained from the redshift distribution given in
Equation~19 of TO09 and a simple luminosity dependence on viewing angle given as 
\begin{equation}
\begin{aligned}
	L(\theta_v) = L_o \; ; \theta_v < \theta_j, \\
	L(\theta_v) = L_o \exp{\left( \frac{(\Gamma\theta_v - \Gamma\theta_j)^2}{2 }
	\right)} \;
	; \theta_v > \theta_j, \\
	F_G = \frac{L(\theta_v)}{4\pi\, D_L^2}.
\end{aligned}
\end{equation}

\begin{figure}[!ht]
	\centering
	\includegraphics{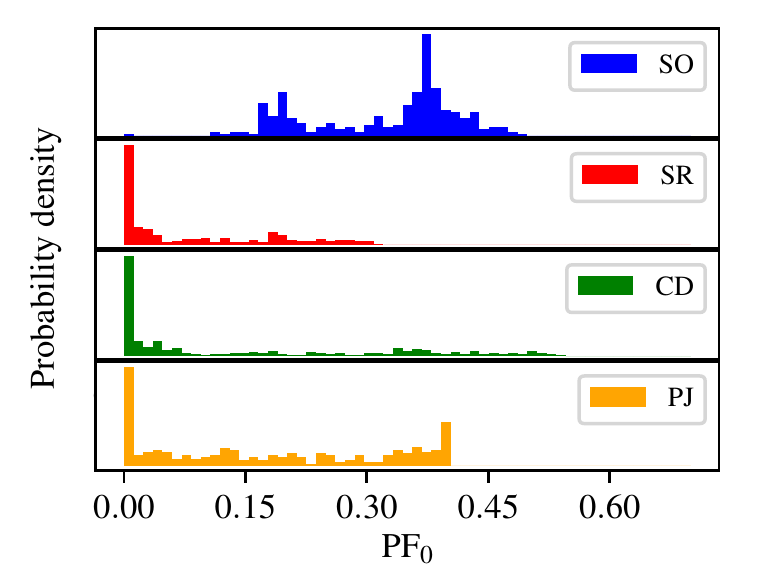}
	\caption{Distribution of the predicted polarisation fraction (PF$_0$) for the four GRB emission models considered: 
	(i) synchrotron emission in an ordered magnetic field (SO), (ii) synchrotron emission in a random magnetic field (SR), (iii) Compton drag (CD), and, (iv) photospheric jet emission (PJ).
	}
	\label{fig:tommod}
\end{figure}

The PF distributions obtained using these assumptions are shown in Fig.~\ref{fig:tommod}.
Synchrotron emission in an ordered magnetic field yields 0.15$ < $PF$ < $0.5. 
For synchrotron emission in a random magnetic field, the distribution is instead 0$ < $PF$ < $0.3.
The Compton drag model favours PF$ < $0.2, with a significant tail to higher values.  
Predictions for the photospheric jet model mirror the Compton drag model for low PF values, with the addition of a second small peak, corresponding to the jet edge, at PF$_\mathrm{max}$.

The statistical properties of the measured PF are given by the Rice distribution,
\begin{equation}
\begin{aligned}
  p(\mathrm{PF},\mathrm{PF}_0,\sigma_\mathrm{PF}) = \\
    \frac{\mathrm{PF}}{\sigma_\mathrm{PF}^2}\exp{\left(-\frac{\mathrm{PF}^2 + \mathrm{PF}_0^2}{2\sigma_\mathrm{PF}^2}\right)}I_0\left( \frac{\mathrm{PF}\times \mathrm{PF}_0}{\sigma_\mathrm{PF}^2} \right),
\end{aligned}
\label{eqn:rice}
\end{equation}
where PF$_0$ is the prediction from the models in Fig.~\ref{fig:tommod}, $\sigma_\mathrm{PF}$ is the uncertainty, and $I_0$ is the modified Bessel function of the zeroth order.
Systematic errors arising from an incorrect estimate of $M_{100}$ due to uncertainties on the GRB energy spectrum or sky position are not considered. 
\begin{figure}[!ht]
	\centering
	\includegraphics{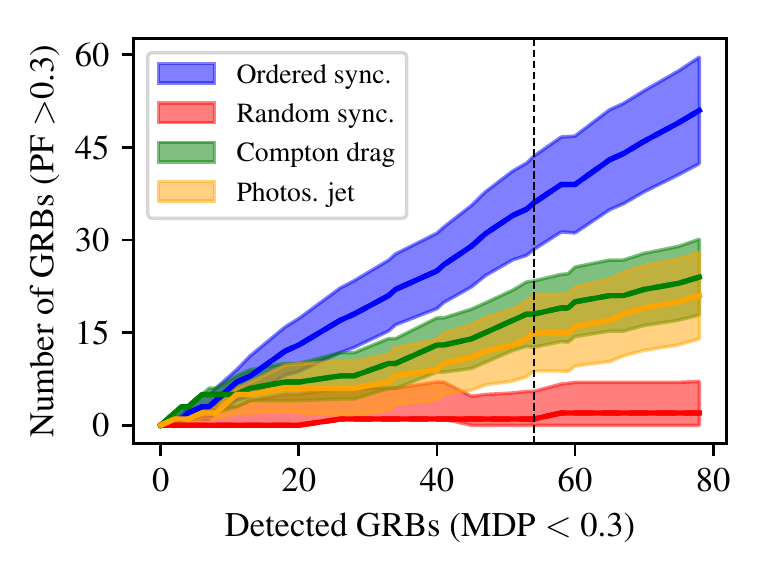}
	\caption{Model separation (at 99.7\% confidence level) as a function of the number of GRBs detected with MDP $<$ 0.3.
	The vertical dashed line marks the number of GRBs expected to be observed by SPHiNX at MDP $<$ 0.3 during a two-year mission.}
	\label{fig:modsep}
\end{figure}

The number of GRBs detected with a PF value exceeding a given threshold value is a straight-forward way to discriminate between emission models, as shown in Fig.~\ref{fig:modsep}
(a more rigorous approach such as a Bayesian model comparison can also be used but is outside of the scope of this paper).
The error bands shown in the figure are obtained assuming that the number of GRBs greater than the PF threshold will be distributed as the sum of individual Bernoulli trials each with the probability given in Equation~\ref{eqn:rice} and a standard deviation, $\sigma_{B}$, given as
\begin{equation}
\sigma_{B} = \sum\limits_i p_i(\mathrm{PF},\mathrm{PF}_0,\sigma_\mathrm{PF})\left( 1 - p_i(\mathrm{PF},\mathrm{PF}_0,\sigma_\mathrm{PF}) \right),  
\end{equation}
where the sum is made over all GRBs exceeding the PF threshold.

Fermi-GBM sees $\sim$250~GRBs per year with an all-sky field-of-view (barring earth occultation).
SPHiNX has greater than double the effective area, and a field-of-view comprising a quarter of the sky. 
Taking into account differences in background rates between the missions, it is estimated that SPHiNX will see $\sim$200~GRBs during the 2~year mission. 
Significant polarisation measurements can be made for $200 \times 532/1967 \simeq 54$ GRBs. With this level of GRB statistics, the synchrotron-based models are manifestly separable.
In order to discriminate between the Compton drag and Photospheric jet models, spectral correlations with PF are required, wherein the photospheric model is expected to show higher PF values for softer spectra, while no such correlation exists for the Compton drag model (see Table \ref{table:correlate}).
%
%
%%%%%%%%%%%%%%%%%%%%%%%%
%
\subsection{Jet structure and magnetisation}
\label{sss:scijet}
SPHiNX can provide simultaneous energy spectrum and light-curve measurements, there-by permitting energy and time-binned polarisation studies for the brightest GRBs. 
Time-resolved variations in PA can be used to probe jet structure as described in Table~\ref{table:correlate}.
To exemplify this, MDP values were derived for all GRBs in the Fermi-GBM catalogue.
Each burst was divided into two equal time bins under the simplified assumption of uniform fluence.
Of the 1967 GRBs, 342 yield \mbox{MDP $<$ 0.3}, there-by indicating that time resolved polarisation studies can be performed for $\sim$35 GRBs during a two year mission.  
It is also instructive to consider the performance of SPHiNX for the time-varying PA claimed by the GAP mission for GRB100826A. 
SPHiNX reconstructs an MDP of $\sim$5\% for the entire burst and $\sim$8\% for each time bin. 
The uncertainty on PA~\cite{Mikhalev18} is $\sim$2.5$^\circ$ ($\sim$20$^\circ$ for GAP) in each time bin. 

The reconstruction of energy-resolved polarisation parameters above and below the Band function peak aids in model comparison, as discussed
in Section~\ref{sec:motivation}. The question of jet magnetisation can readily be answered once the emission mechanism model is determined.

%%%%%%%%%%%%%%%%%%%%%%%%
%
\subsection{Comparison with AstroSat}
\label{sec:comp}
The projected polarimetric performance of SPHiNX is compared to measurements reported by the on-going AstroSat mission~\cite{astrosatGRB}.
Simulations adopt the set-up described in Section~\ref{sec:bkgnd} with GRB parameters taken from the AstroSat catalogue.
For AstroSat observations of GRBs located in the lower hemisphere, the incidence angle is mirrored into the upper hemisphere.
%
\begin{comment}
\[ \begin{aligned}
	\theta = \theta_{ast} \; ; \theta_{ast} < 90^\circ \\
	\theta = \pi/2 - \theta_{ast} \,(\bmod\,90)  \; ; \theta_{ast} > 90^\circ 
\end{aligned}\]
\end{comment}

\begin{figure}[!ht]
	\centering
	\includegraphics{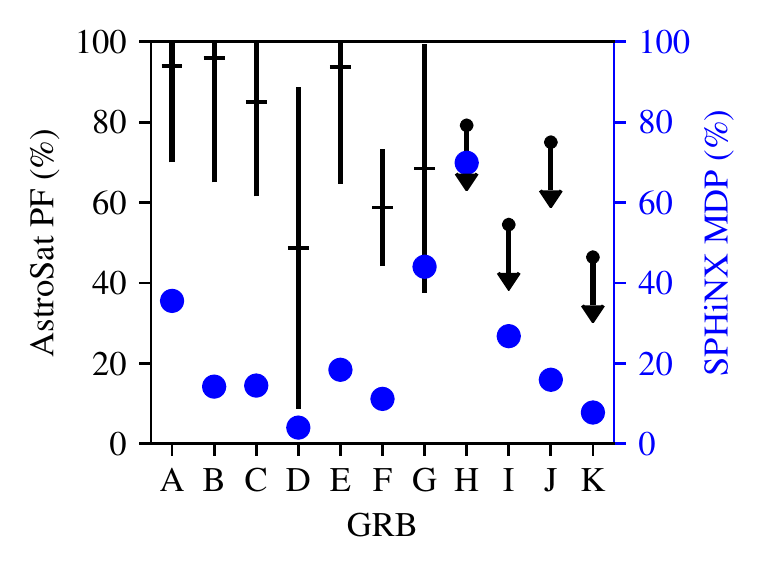}
	\caption{
	A comparison between the PF values reported by AstroSat (black) and
	expected SPHiNX MDP values (blue).
	}
	\label{fig:spas}
\end{figure}

As shown in Table~\ref{tab:spvas} for each GRB reported by AstroSat, the SPHiNX detection significance (obtained by converting the chance
probability of detecting the given signal into standard deviations) has been estimated and is compared to that reported by AstroSat.
For all GRBs, SPHiNX has significantly better performance. 
The SPHiNX MDP value for each GRB is shown in Fig.~\ref{fig:spas} along with the AstroSat PF or upper limit.
\begin{table}
\caption{The expected polarimetric performance of SPHiNX for GRB measurements reported by AstroSat.
For bursts H--K, a 95\% upper limit (UL) is provided by AstroSat.}
\begin{center}
\begin{tabular}{|c|c|c|c|c|}
\hline
{ } & {\bf GRB} & {\bf \makecell{AstroSat \\ (N$\sigma$)}}  & 
{\bf \makecell{SPHiNX \\ (N$\sigma$)}} & {\bf
\makecell{SPHiNX \\ MDP (\%)}}\\
\hline
A & 160131A & $>3.48$ &  $4.42$ & $35.5$\\
B & 160509A & $>3.48$ &  $>10$ & $14.2$\\
C & 160802A & $>3.48$ &  $>10$&  $14.4$\\
D & 160821A & $>3.48$  & $>10$&  $3.9$\\
E & 160910A & $>3.48$ & $7.45$&  $18.4$ \\
F & 160325A & $2.24$ &  $>10$ & $11.1$\\
G & 160106A & $2.36$ & 4.48& $44.0$\\\hline
{ } & {\bf GRB} & {\bf \makecell{AstroSat \\ 95\% UL}}  & 
{\bf \makecell{SPHiNX \\ 95\% UL}} & {\bf  \makecell{SPHiNX \\ MDP (\%)}}\\
\hline
H & 151006A & $79.2$ & 56.4& $69.9$\\
I & 160703A & $54.5$ & $21.6$&  $26.7$\\
J & 160607A & $75$ & $12.8$ & $15.9$\\
K & 160623A & $46.4$ & $6.3$& $7.7$\\\hline
%{ } & {\bf GRB} & {\bf \makecell{GAP \\ (N$\sigma$)}}  & 
%{\bf \makecell{SPHiNX \\ (N$\sigma$)}} & {\bf
%\makecell{SPHiNX \\ MDP (\%)*}}\\
%\hline
%L & 110301A & 3.7 & $>10$ & 9.4 \\
%M & 110721A & 3.3 & $>10$ & 9.2 \\
%N-I & 100826A & 2.29 & $>10$ & 4.5 \\
%N-II & 100826A & 1.92 & $>10$ & 6.6 \\
%\hline 
\end{tabular}
\end{center}
\label{tab:spvas}
\end{table}
%% NEW

%%%%%%%%%%%%%%%%%%%%%%%%%%%%%%%%%%%%%%%%%%%%%%
\section{Outlook}
\label{sec:outlook}

This paper demonstrates that a relatively low cost small satellite mission has the potential to make a significant advance in understanding the properties of GRB jets, and to identify the mechanism behind the high-energy emission.
The use of a dedicated instrument, the polarimetric response of which is fully characterised prior to launch, ensures reliable results.
While the SPHiNX mission proposal was specifically developed for the Swedish InnoSat programme, the design concept is well suited to other small satellite or space-station platforms.
The scientific performance of SPHiNX would be improved if the InnoSat platform could operate in a lower inclination orbit.
This would increase the observation duty-cycle by reducing the number of passages through the SAA. Moreover, background levels (prompt and delayed) would decrease, with corresponding improvement of the MDP.
The first two InnoSat missions, with launches foreseen in 2019 and 2022, concern atmospheric science. The call for the third InnoSat mission is expected late in 2019.

%%%%%%%%%%%%%%%%%%%%%%%%%%%%%%%%%%%%%%%%%%%%%%
\section{Acknowledgements}

SPHiNX activities at KTH were funded by The Swedish National Space Agency and the KTH Space Centre. 
L.\,Eriksson, S.\,Chandrasekhar, C.\,Fuglesang, S.\,Grahn, N.\,Ivchenko, K.\,Lindwall, G.\,Olentsenko, G.\,Tibert, and N.\,Uchida are thanked for their contributions to the Phase A studies.
Contributions to the development of the SPHiNX mission concept by past KTH team members are gratefully acknowledged.
OHB Sweden are thanked for an efficient collaboration and for providing technical details of the InnoSat platform.

%%%%%%%%%%%%%%%%%%%%%%%%%%%%%%%%%%%%%%%%%%%%%%%%%%%%%%%%
%
\bibliographystyle{elsarticle-num}

%%%%%%%%%%%%%%%%%%%%%%%%%%%%%%%%%%%%%%%%%%%%%%%%%%%%%%%%

\end{document}